\documentclass[12pt]{article}
\usepackage{latexsym}
\usepackage{amsmath}
\usepackage{amssymb}

\def \bbe {\mbox{\boldmath $e$}}
\def \bbf {\mbox{\boldmath $f$}}
\def \bbg {\mbox{\boldmath $g$}}

\def \bbu {\mbox{\boldmath $u$}}
\def \bbv {\mbox{\boldmath $v$}}
\def \bbw {\mbox{\boldmath $w$}}

\def \bvphi {\mbox{\boldmath $\varphi$}}

\newcommand{\IR}{{\Bbb R}}

\newcommand{\IC}{{\Bbb C}}

\newcommand{\clb}{{\cal B}} 
\newcommand{\clc}{{\cal C}}
\newcommand{\cld}{{\cal D}}
\newcommand{\cle}{{\cal E}} 

\newcommand{\clh}{{\cal H}}

\def \qed {\hfill \vrule height6pt width 6pt depth 0pt}

\begin{document}
\thispagestyle{empty}
\centerline{\Large{\bf Extremal Quantum States in Coupled Systems}}
\centerline{\bf by}
\centerline{\bf K. R. Parthasarathy}
\centerline{\bf Indian Statistical Institute, Delhi Centre,}
\centerline{\bf 7, S. J. S. Sansanwal Marg, }
\centerline{\bf New Delhi - 110 016, India.}
\centerline{\bf e-mail : krp@isid.ac.in}
\vskip40pt
\centerline{In memory of Paul Andr\'e Meyer}
\vskip1.5in
\begin{abstract}
Let ${\cal H}_1,$ ${\cal H}_2$ be finite dimensional complex Hilbert spaces describing the states of two finite level quantum systems. Suppose $\rho_i$ is a state in ${\cal H}_i, \:\:i=1,2.$ Let ${\cal C} (\rho_1, \rho_2)$ be the convex set of all states $\rho$ in ${\cal H} = {\cal H}_1 \otimes {\cal H}_2$ whose marginal states in ${\cal H}_1$ and ${\cal H}_2$ are $\rho_1$ and $\rho_2$ respectively. Here we present a necessary and sufficient criterion for a $\rho$ in ${\cal C} (\rho_1, \rho_2)$ to be an extreme point. Such a condition implies, in particular, that for a state $\rho$ to be an extreme point of ${\cal C} (\rho_1, \rho_2)$ it is necessary that the rank of $\rho$ does not exceed $\left (d_1^2 + d_2^2 - 1 \right )^{\frac{1}{2}},$ where $d_i = \: \dim\:{\cal H}_i, \:\:i=1,2.$ When ${\cal H}_1$ and ${\cal H}_2$ coincide with the $1$-qubit Hilbert space $\IC^2$ with its standard orthonormal basis $\left \{ |0 >, |1> \right \}$ and $\rho_1 = \rho_2 = \frac{1}{2} I$ it turns out that a state $\rho \in {\cal C} (\frac{1}{2}I, \frac{1}{2}I)$ is extremal if and only if $\rho$ is of the form $|\Omega>< \Omega|$ where $| \Omega > = \frac{1}{\sqrt{2}} \left (|0> | \psi_0 > + |1 > | \psi_1 > \right ),$ $\left \{ | \psi_0 >, | \psi_1>    \right \}$ being an arbitrary orthonormal basis of $\IC^2.$ In particular, the extremal states are the maximally entangled states.

\end{abstract}

\newpage
\setcounter{page}{1}
\section{Introduction} One of the well-known problems of classical probability theory is the determination of the set of all extreme points in the convex set of all probability distributions in a product Borel space $\left (X \times Y, \:\:{\cal F} \times {\cal G} \right )$ with fixed marginal distributions $\mu$ and $\nu$ on $\left (X, {\cal F} \right )$ and $\left (Y, {\cal G} \right )$ respectively. Denote this convex set by $C(\mu, \nu).$ When $X = Y = \left \{ 1,2,\ldots,n \right \},$ ${\cal F} = {\cal G}$ is the field of all subsets of $X$ and $\mu = \nu$ is the uniform distribution then the problem is answered by the famous theorem of Birkhoff [1] that the set of extreme points of the convex set of all doubly stochastic matrices of order $n$ is the set of all permutation matrices of order $n.$ Problems of this kind have a natural analogue in quantum probability. Suppose ${\cal H}_1$ and ${\cal H}_2$ are finite dimensional complex Hilbert spaces describing the states of two finite level quantum systems $S_1$ and $S_2$ respectively. Then the Hilbert space of the coupled system $S_{12}$ is ${\cal H}_1 \otimes {\cal H}_2.$ Suppose $\rho_i$ is a state of $S_i$ in ${\cal H}_i, \: \: i=1,2.$ Any state $\rho$ in $S_{12}$ yields marginal states $\mbox{Tr}_{{\cal H}_2} \rho$ in ${\cal H}_1$ and $\mbox{Tr}_{{\cal H}_1} \rho$ in ${\cal H}_2$ where $\mbox{Tr}_{{\cal H}_i} $ is the relative trace over ${\cal H}_i.$ Denote by ${\cal C} \left (\rho_1, \rho_2 \right )$ the convex set of all states $\rho$ of the coupled system $S_{12}$ whose marginal states  in ${\cal H}_1$ and ${\cal H}_2$ are $\rho_1$ and $\rho_2$ respectively. One would like to have a complete description of the set of all extreme points of ${\cal C} \left (\rho_1, \rho_2 \right ).$ In this paper we shall present a necessary and sufficient criterion for an element $\rho$ in ${\cal C} \left (\rho_1, \rho_2 \right ) $ to be an extreme point. This leads to an interesting (and perhaps surprising) upper bound on the rank of such an extremal state $\rho.$ Indeed, if $\rho$ is an extreme point of ${\cal C} \left (\rho_1, \rho_2 \right ) $ then the rank of $\rho$ cannot exceed $\left (d_1^2 + d_2^2 - 1  \right )^{\frac{1}{2}}$  where $d_i = \dim \: {\cal H}_i.$ Note that the rank of an arbitrary state in ${\cal H}_1 \otimes {\cal H}_2 $ can vary from 1 to $d_1d_2.$ When ${\cal H}_1 = {\cal H}_2 = \IC^2,$  $\left \{|0>, |1>\right \}$ is the standard (computational) basis of $\IC^2$ and $\rho_1 = \rho_2 = \frac{1}{2}I$ it turns out that a state $\rho$ in ${\cal C} \left (\frac{1}{2}I, \frac{1}{2} I \right )$ is extremal if and only if $\rho$ has the form $|\Omega><\Omega|$ where $|\Omega>=\frac{1}{\sqrt{2}} \left (|0>|\psi_0> + |1>|\psi_1> \right ),$ $\left \{|\psi_0>, |\psi_1 >   \right \}$ being any orthonormal basis of $\IC^2.$ These are the well-known maximally entangled states.

This work was done  by the author during his visit to the University of Greifswald during 17 June - 16 July under a DST (India) - DAAD (Germany) project between the Indian Statistical Institute and the mathematics department of the University of Greifswald. The author is grateful to these organisations for their generous support. The hospitality extended by the colleagues of the Quantum Probability group in the University of Greifswald and, particularly, Michael Schurmann is gratefully acknowledged.

\section{Extreme points of the convex set ${\cal C} \left ( \rho_1, \rho_2 \right )$} In the analysis of extreme points in a compact convex set of positive definite matrices the following proposition plays an important role [5]. See also [2-4].

\vskip20pt
\noindent {\bf Proposition 2.1} Let $\rho$ be any positive definite matrix of order $n$ and rank $k < n.$ Then there exists a permutation matrix $\sigma$ of order $n,$ a $k \times( n-k)$ matrix $A$ and a strictly positive definite matrix $K$ of order $k$ such that

\begin{equation}
\sigma \rho \sigma^{-1} = \left [\begin{array}{c|c} K & KA \\ \hline  A^{\dagger} K &  A^{\dagger}KA \end{array}   \right ] \label{a1}
\end{equation}
If, in addition, $\rho = \frac{1}{2} \left (\rho^{\prime} + \rho^{\prime \prime} \right )$ where $\rho^{\prime}$ and $\rho^{\prime \prime}$ are also positive definite matrices then there exist positive definite matrices $K^{\prime},$ $K^{\prime \prime}$ of order $k$ such that

\begin{equation}
\sigma \rho^{\#} \sigma^{-1} = \left [\begin{array}{c|c} K^{\#} & K^{\#} A \\ \hline  A^{\dagger} K^{\#} &  A^{\dagger}K^{\#}A \end{array}   \right ] \label{a2}
\end{equation}
where $\#$ indicates $\prime$ and $\prime \prime.$
\vskip10pt
\noindent {\bf Proof:} Choose vectors $\bbu_ i  \in \IC^n, \:\: i=1,2,\ldots, n$ such that
$$\rho = \left ( \left ( \langle \bbu_i |  \bbu_j \rangle   \right ) \right ), \:\: i, j \in \left \{1,2,\ldots, n  \right \}.  $$
Since rank $\rho = k,$ the linear span of all the $\bbu_i$'s has dimension $k.$ Hence modulo a permutation $\sigma$ of $\left \{1,2, \ldots, n \right \}$ we may assume that $\bbu_1, \bbu_2, \ldots, \bbu_k$ are linearly independent and 
\begin{equation}
\bbu_{k+j} = a_{1j} \bbu_1 + a_{2j} \bbu_2 + \cdots + a_{kj} \bbu_k, \:\: 1 \leq j \leq n - k.    \label{a3}
\end{equation}
Putting
\begin{eqnarray*}
K &=& ((\langle \bbu_i | \bbu_j \rangle )), \:\:i,j \in {1,2,\ldots, k}, \\
A &=& ((a_{ij})), \:\: i=1,2,\ldots, k; \:\: j = 1,2,\ldots,n-k
\end{eqnarray*}
and denoting by the same letter $\sigma,$ the permutation unitary matrix of order $n$ corresponding to $\sigma$ we obtain the relation (\ref{a1}). To prove the second part we express
$$\sigma \rho \sigma^{-1} = \left [\begin{array}{c|c} K & KA \\ \hline  A^{\dagger} K &  A^{\dagger}KA \end{array} \right ]  = \frac{1}{2}   \left [\begin{array}{c|c} K^{\prime} & B_1 \\ \hline  B_1^{\dagger} &  C_1 \end{array} \right ] + \frac{1}{2} \left [\begin{array}{c|c} K^{\prime \prime} & B_2 \\ \hline  B_2^{\dagger}  &  C_2 \end{array} \right ]  $$ 
where the two partitioned matrices on the right hand side are the matrices $\sigma \:\rho^{\prime} \sigma^{-1}$ and $\sigma \rho^{\prime \prime} \sigma^{-1}.$ Now construct vectors ${\bbv_i},$ ${\bbw_i},$ $ i = 1,2,\ldots, n$ such that
\begin{eqnarray}
\sigma \rho^{\prime} \sigma^{-1} &=& (( \langle \bbv_i | \bbv _j \rangle)), \:\: i,j \in \{1,2,\ldots, n\}   \label{a4} \\
\sigma \rho^{\prime \prime} \sigma^{-1} &=& ((\langle \bbw_i | \bbw _j   \rangle)), \:\: i, j \in \{1,2, \ldots, n\} . \label{a5}
\end{eqnarray}
Let ${|0>, |1>}$ be the standard orthonormal basis of $\IC^2.$ Define
\begin{equation}
| \bvphi_i > = \frac{1}{\sqrt{2}} (|\bbv_i> |0> + | \bbw_i> |1>), \:\:1 \leq i \leq n.  \label{a6}
\end{equation}
Then we have
\begin{eqnarray*}
<\bvphi_i | \bvphi_j> &=& \frac{1}{2} (\langle \bbv _i | \bbv _j \rangle + \langle \bbw _i |  \bbw_j)\\
&=& \langle \bbu_i | \bbu_j \rangle \quad \mbox{for all}\:\: i, j  \{1,2, \ldots, n\}.
\end{eqnarray*}
Thus the correspondence $\bbu_i \rightarrow \bvphi_i$ is an isometry. Hence by (\ref{a3}) we have
$$\bvphi_{k+j} = a_{1j} \bvphi_1 + a_{2j}  \bvphi_2 + \cdots + a_{kj} \bvphi_k, \:\: 1 \leq j \leq n-k.$$
Substituting for the $\bvphi_i$'s from (\ref{a6}) and using the orthogonality of $|0>$ and $|1>$ we conclude that
\begin{eqnarray}
| \bbv_{k+j}>&=& \sum_{i=1}^k a_{ij} | \bbv_i >, \label{a7} \\
|\bbw_{k+j}>&=& \sum _{i=1}^k a _{ij} | \bbw _i >. \label{a8}
\end{eqnarray}
Putting
\begin{eqnarray*}
K^{\prime} &=& ((\langle \bbv _i | \bbv_j \rangle)), \quad i, j \in \{1, 2, \ldots, k\}\\
K^{\prime \prime} &=& ((\langle \bbw_i | \bbw_j \rangle)), \quad i, j \in =\{1,2,\ldots, k\}
\end{eqnarray*}
and substituting (\ref{a7}) and (\ref{a8}) in (\ref{a4}) and (\ref{a5}) we obtain $B_1 = K^{\prime} A,$ $C_1 = A^{\dagger} K^{\prime} A,$ $B_2 = K^{\prime \prime} A,$ $C_2 = A^{\dagger} K^{\prime \prime} A.$ Thus we have (\ref{a2}). \qed

Let $\clh_1,$ $\clh_2$ be two complex Hilbert spaces of finite dimension $d_1, d_2$ and equipped with orthonormal bases $\{\bbe_1, \bbe_2, \ldots, \bbe_{d_{1}} \}$, $\{\bbf_1, \bbf_2, \ldots, \bbf_{d_{2}}\}$ respectively. Consider the tensor product $\clh = \clh_1 \otimes \clh_2$ equipped with the orthonormal basis $\bbg _{ij} = \bbe _i \otimes \bbf _j$ with the ordered pairs $ij$ in the lexicographic order. For any operator $X$ on $\clh$ we associate its marginal operators $X_i$ in $\clh_i$ by putting
$$X_1 = \mbox{Tr}_{\clh_{2}} X, \quad X_2 = \mbox{Tr}_{\clh_{1}} X $$
where $\mbox{Tr}_{\clh_{i}}$ stands for the relative trace over $\clh_i.$ If $\rho$ is a state on $\clh,$ i.e., a positive operator of unit trace, then its marginal operators are states in $\clh_1$ and $\clh_2.$ Now we fix two states $\rho_1$ and $\rho_2$ in $\clh_1$ and $\clh_2$ respectively and consider the compact convex set
$$\clc (\rho_1, \rho_2) = \{\rho | \rho \:\:\mbox{a state on}\: \clh  \:\: \mbox{with marginals}  \:\:\rho_1 \:\: \mbox{and} \:\: \rho_2  \:\:\mbox{in}  \:\:\clh_1  \:\:\mbox{and}  \:\:\clh_2 \:\: \mbox{respectively. } \} $$
in $\clb (\clh).$ Let $\cle (\rho_1, \rho_2) \subset \clc (\rho_1, \rho_2)$ be the set of all extreme points in $\clc (\rho_1, \rho_2).$

\vskip10pt
\noindent {\bf Proposition 2.2} Let $\rho \in \cle (\rho_1, \rho_2).$ Then $\rho$ is singular.
\vskip 10pt
\noindent {\bf Proof:} Suppose $\rho$ is nonsingular. Choose nonzero hermitian operators $L_i$ in $\clh_i$ with zero trace. Then for all sufficiently small and positive $\varepsilon,$ the operators $\rho \pm \varepsilon L_1 \otimes L_2$ are positive definite. Since the marginal operators of $L_1 \otimes L_2$ are $0,$ both of the operators $\rho \pm \varepsilon L_1 \otimes L_2$ belong to $\clc (\rho_1, \rho_2)$ and
$$\rho = \frac{1}{2} \left ((\rho + \varepsilon L_1 \otimes L_2) + (\rho - \varepsilon L_1 \otimes L_2) \right ) $$
and $\rho$ is not extremal. \qed
\vskip20pt
\noindent {\bf Proposition 2.3} Let $n = d_1 d_2, \quad \rho \in \clc (\rho_1, \rho_2),$ rank $\rho = k < n$ and let $\sigma$ be a permutation of the ordered basis $\{\bbg_{ij}\}$ of $\clh$ such that
\begin{equation}
\sigma \rho \sigma^{-1} = \left [\begin{array}{c|c} K & KA \\ \hline A^{\dagger}K & A^{\dagger}KA \end{array}  \right ], \label{a9}
\end{equation}
where $K$ is a strictly positive definite matrix of order $k.$ Then, in order that $\rho  \in \cle (\rho_1, \rho_2)$ it is necessary that there exists no nonzero hermitian matrix $L$ of order $k$ such that both the marginal operators of
\begin{equation}
\sigma^{-1} \left [\begin{array}{c|c}  L & LA \\ \hline A^{\dagger}L & A^{\dagger}LA \end{array} \right ] \sigma   \label{a10}
\end{equation}
vanish.
\vskip20pt
\noindent {\bf Proof:} Suppose there exists a nonzero hermitian matrix $L$ of order $k$ such that both the marginals of the operator (\ref{a10}) vanish. Since $K$ in (\ref{a9}) is nonsingular and positive definite it follows that for all sufficiently small and positive $\varepsilon,$ the matrices $K \pm \varepsilon\: L$ are strictly positive definite. Hence
$$\rho = \frac{1}{2} \left \{\sigma^{-1} \left [\begin{array}{c|c}K + \varepsilon L & (K+\varepsilon L)A \\ \hline A^{\dagger} (K+\varepsilon L) & A^{\dagger} (K+\varepsilon L)A   \end{array}  \right ] \sigma +  \sigma^{-1} \left [\begin{array}{c|c}K - \varepsilon L & (K-\varepsilon L)A \\ \hline A^{\dagger} (K-\varepsilon L) & A^{\dagger} (K-\varepsilon L)A   \end{array}  \right ] \sigma \right \} $$
where each summand on the right hand side has the same marginal operators as $\rho.$ Furthermore
$$\left [ \begin{array}{c|c}K \pm \varepsilon L & (K \pm \varepsilon L) \\ \hline  A^{\dagger} (K \pm \varepsilon L) &   A^{\dagger}(K \pm \varepsilon L)A  \end{array}  \right ] = \left [\frac{I}{A^{\dagger}} \right ] \left (K \pm \varepsilon L \right ) [I|A] \ge 0. $$
Thus $\rho$ is not extremal. \qed
\vskip20pt

\noindent {\bf Corollary} Let $\rho \in \cle (\rho_1, \rho_2).$ Then rank $\rho \leq \sqrt{d_1^2 + d_2^2 - 1}.$
\vskip20pt
\noindent {\bf Proof:} Let rank $\rho = k.$ By proposition 2.2, $k<n.$ Since $\rho$ is a positive definite matrix in the basis $\{\bbg_{ij}\}$ such that  $\sigma \rho \sigma^{-1} $ can be expressed in the form (\ref{a9}). The extremality of $\rho$ implies that there exists no nonzero hermitian matrix $L$ of order $k$ such that the matrix (\ref{a10}) has both its marginals equal to $0.$ The vanishing of both the marginals of (\ref{a10}) is equivalent to
\begin{equation}
\mbox{Tr}\: \sigma^{-1} \left [\begin{array}{c|c}L & LA \\ \hline A^{\dagger} L & A^{\dagger} LA    \end{array}  \right ] \sigma \left ( X_1 \otimes I^{(2)} + I^{(1)} \otimes X_2 \right ) = 0    \label{a11}
\end{equation}
for all hermitian operators $X_i$ in $\clh_i,$ $I^{(i)}$ being the identity operator in $\clh_i.$ Equation (\ref{a11}) can be expressed as 
$$\mbox{Tr}\: L \: \left [ I_k | A \right ] \sigma \left (X_1 \otimes I^{(2)} + I^{(1)} \otimes X_2 \right ) \sigma^{-1} \left [\frac{I_k}{A^{\dagger}} \right ] = 0. $$
In other words $L$ is in the orthogonal complement of the real linear space
$$\cld = \left \{\left . [I_k | A] \sigma \left (X_1 \otimes I^{(2)} + I^{(1)} \otimes X_2 \right ) \sigma ^{-1} \left [\frac{I_k}{A^{t}} \right ] \right | X_i \:\: \mbox{hermitian in}\: \clh_i, i=1,2 \right \} , $$
with respect to the scalar product $\langle L | M \rangle = \:\mbox{Tr}\: LM$ between any two hermitian matrices of order $k.$ Thus the extremality of $\rho$ implies that $\cld ^{\perp} = \{0\}.$ The real linear space of all hermitian matrices of order $k$ has dimension $k^2.$ The real linear space of all hermitian operators of the form $X_1 \otimes I^{(2)} + I^{(1)} \otimes X_2$  is $d_1^2 + d_2^2-1.$ Thus $k^2 = \dim \cld \leq d_1^2 + d_2^2-1.$ \qed

\vskip20pt
\noindent {\bf Proposition 2.4} Let $\rho \in \clc \left (\rho_1, \rho_2 \right ), k, \sigma, K, A$ be as in Proposition 2.3. Suppose there is no nonzero hermitian matrix $L$ of order $k$ such that both the marginal operators of
$$\sigma^{-1} \left [\begin{array}{c|c}L & LA \\ \hline A^{\dagger} L & A^{\dagger} LA  \end{array}  \right ] \sigma  $$ 
vanish. Then $\rho \in \cle (\rho_1, \rho_2).$
\vskip10pt
\noindent {\bf Proof:} Suppose $\rho {\not \in} \cle (\rho_1, \rho_2).$  Then there exist two distinct states $\rho^{\prime}, \rho^{\prime \prime}$ in $\clc (\rho_1, \rho_2)$ such that
$$\rho = \frac{1}{2}  (\rho ^{\prime} + \rho^{\prime \prime}), \quad \rho^{\prime} \neq \rho^{\prime \prime }.$$
Since rank $\rho = k$ it follows from Proposition 2.1 that there exist positive definite matrices $K^{\prime},$ $K^{\prime \prime}$ of order $k$ such that
$$\sigma \rho^{\#} \sigma^{-1} = \left [\begin{array}{c|c} K^{\#} & K^{\#} A \\ \hline A^{\dagger} K^{\#} & A^{\dagger} K^{\#} A \end{array}  \right ]$$ 
where $\left (\rho^{\#}, K^{\#}  \right )$ stands for any of the three pairs $(\rho, K), $ $(\rho^{\prime}, K^{\prime}),$ $(\rho^{\prime \prime}, K^{\prime \prime}).$ Since $\rho^{\prime} \neq \rho^{\prime \prime}$ and hence $\sigma  \rho^{\prime } \sigma^{-1} \neq \sigma \rho^{\prime \prime} \sigma^{-1}$ it follows that $K^{\prime} \neq K^{\prime \prime}.$ Putting $L = K^{\prime} - K^{\prime \prime} \neq 0$ we obtain a nonzero hermitian matrix $L$ of order $k$ such that both the marginal operators of
$$ \sigma^{-1} \left [\begin{array}{c|c}L & LA \\ \hline A^{\dagger}L & A^{\dagger} LA   \end{array} \right ] \sigma $$
vanish. This is a contradicton. \qed

Combining Proposition 2.3, its Corollary and Proposition 2.4 we have the following theorem.
\vskip10pt
\noindent {\bf Theorem 2.5} Let $\clh_1,$ $\clh_2$ be complex finite dimensional Hilbert spaces of dimension $d_1,$ $d_2$ respectively. Suppose $\clc (\rho_1, \rho_2)$ is the convex set of all states $\rho$ in $\clh = \clh_1 \otimes \clh_2$ whose marginal states in $\clh_1$ and $\clh_2$ are $\rho_1$ and $\rho_2$ respectively. Let $\{\bbe_{i}\}$, $\{\bbf_{j} \}$ be orthonormal bases for $\clh_1,$ $\clh_2$ respectively and let $\bbg _{ij} = \bbe_i \otimes \bbf_j,$ $i=1,2,\ldots, d_1;$ $j = 1,2, \ldots, d_2$ be the orthonormal basis of $\clh$ in the lexicographic ordering of the ordered pairs $ij.$ In order that an element $\rho$ in $\clc (\rho_1, \rho_2)$ be  an extreme point it is necessary that its rank $k$ does not exceed $\sqrt{d_1^2 + d_2^2 -1}.$ Let $\sigma$ be a permutation unitary operator in $\clh,$ permuting the basis $\{\bbg _{ij}\}$ and satisfying
$$\sigma \rho \sigma^{-1} = \left [\begin{array}{c|c}K & KA \\ \hline A^{\dagger} K & A^{\dagger} KA   \end{array}  \right ]  $$
where $K$ is a strictly positive definite matrix of order $k.$ Then $\rho$ is an extreme point of the convex set $\clc (\rho_1, \rho_2)$ if and  only if the real linear space
$$\cld = \left \{\left . [I_k | A]  \sigma \left (X_1 \otimes I^{(2)} + I^{(1)} \otimes X_2 \right ) \:\sigma^{-1}\: \left [\frac{I}{A^{t}} \right ]  \right  | X_i \:\mbox{hermitian in } \:\clh_i, \:i=1,2\right \} $$
coincides with the space of all hermitian matrices of order $k.$
\vskip10pt
\noindent {\bf Proof: } Immediate from Proposition 2.3, its Corollary and Proposition 2.4. \qed

\section{The case $\clh_1 = \clh_2 = \IC^2$}
\setcounter{equation}{0}
 We consider the orthonormal basis
$$| 0 > =\left  [\begin{array}{c} 1 \\ 0 \end{array}\right ], \quad | 1> = \left [\begin{array}{c} 0 \\ 1 \end{array}\right ]$$
in $\IC^2$ and write
$$| xy > = |x> \otimes |y> \:\mbox{for all}\: x, y \in \{0,1\} .$$
Then $\bbe_1 = |00>,$ $\bbe_2 = |01>,$ $\bbe_3 = |10>,$ $\bbe_4 = |11>$ constitute an ordered orthonormal basis for $\IC^2 \otimes \IC^2.$ For any state $\rho$ in $\IC^2 \otimes \IC^2$ define
\begin{equation}
K_{\rho} \left ((x,y), (x^{\prime}, y^{\prime}) \right ) = \langle xy | \rho | x^{\prime} y^{\prime} \rangle \: \: x, y, x^{\prime}, y^{\prime} \in \{0,1\} .  \label{b1}
\end{equation}
If $\rho$ has marginal states $\rho_1,$   $ \rho_2$ then
\begin{eqnarray}
K_{\rho} \left ((x,0), (x^{\prime}, 0) \right ) &+&  K_{\rho} \left ((x,1), (x^{\prime},1) \right ) = \langle x | \rho_1 | x^{\prime}  \rangle , \label{b2}\\
K_{\rho} \left ((0, y), (0, y^{\prime}) \right ) &+&  K_{\rho} \left ((1,y), (1, y^{\prime}) \right ) = \langle y|\rho_2| y^{\prime}  \rangle   \label{b3}
\end{eqnarray}
for all $x,y,x^{\prime}, y^{\prime}$ in $\{0,1\}.$ If $\rho$ is an extreme point of the convex set $\clc (\rho_1, \rho_2)$ it follows from Theorem 2.5 that the rank of  $\rho$ cannot exceed $\sqrt{7}.$ In other words, every extremal state $\rho^{\prime}$ in $\clc (\rho_1, \rho_2)$ has rank 1 or 2. When $\rho_1 = \rho_2 = \frac{1}{2} I $ we have the following theorem :
\vskip20pt
\noindent {\bf Theorem 3.1} Let $\clh_1 = \clh_2 = \IC^2.$ A state $\rho$ in $\clc (\frac{1}{2} I, \frac{1}{2}I)$ is an extreme point if and only if $\rho = |\Omega >< \Omega |$ where 
$$| \Omega > = \frac{1}{\sqrt{2}} \left (| 0 > \otimes | \psi_0  > + |1> \otimes |\psi_1>\right ), $$
$\{|\psi_0 >, |\psi_1> \}$ being an orthonormal basis of $\IC^2.$
\vskip20pt
\noindent {\bf Proof:} We shall first show that there is no extremal state $\rho$ of rank 2 in $\clc (\frac{1}{2}I, \frac{1}{2} I).$ To this end choose and fix a state $\rho$ of rank 2 in $\clc (\frac{1}{2}I, \frac{1}{2}I).$ Then the right hand sides of (\ref{b2}) and (\ref{b3}) coincide with $\frac{1}{2} \delta_{xx^{\prime}}$ and $\frac{1}{2} \delta_{yy^{\prime}}$ respectively and in the ordered basis $\{\bbe_j, \:1 \leq j \leq 4\}$ the positive definite matrix $K_{\rho}$ of rank 2 in (\ref{b1}) assumes the form

\begin{equation}
K_{\rho} = \left [\begin{array}{cccc} \frac{a}{2} & x & y & z \\ \bar{x} & \frac{1-a}{2} & t & -y \\ \bar{y} & \bar{t} & \frac{1-a}{2} & -x \\ \bar{z} & -\bar{y} & -\bar{x} & \frac{a}{2}   \end{array}  \right ] \label{b4}
\end{equation}
for some $0 \leq a \leq 1,$ $x,y,z,t \in \IC.$ The fact $K_{\rho}$ has rank 2 implies that one of the following three cases holds :
\begin{itemize}
\item [(1)] $\left [\begin{array}{cc} \frac{a}{2} & x \\ \bar{x} & \frac{1-a}{2} \end{array} \right ]$ is strictly positive definite ;

\item [(2)] $\left [\begin{array}{cc} \frac{a}{2} & y \\ \bar{y} & \frac{1-a}{2} \end{array} \right ]$ is strictly positive definite ;

\item [(3)] $|x|^2 = |y|^2 = \frac{a(1-a)}{4}$ and one of the matrices $\left [\begin{array}{cc} \frac{a}{2} & z \\ \bar{x} & \frac{a}{2}   \end{array} \right ]$,  $\left [\begin{array}{cc} \frac{1-a}{2} & t \\ \bar{t} & \frac{1-a}{2}   \end{array} \right ]$
is strictly positive definite.
\end{itemize}

We shall  first show that case (3) is vacuous. We assume that
\begin{equation}
|x|^2 = |y|^2 = \frac{a(1-a)}{4}, \:|z|^2 < \frac{a^2}{4}, \quad \mbox{rank}\: K_{\rho} = 2. \label{b5}
\end{equation}
conjugation by the unitary permutation matrix corresponding to the permutation (1)(24)(3) brings (\ref{b4}) to the form

\begin{equation}
\left [ \begin{array}{c|c}\begin{array}{cc} \frac{a}{2} & z \\ \bar{z} & \frac{a}{2}  \end{array} &   \begin{array}{cc} y & x \\  -\bar{x} & - \bar{y} \end{array} \\ \hline \begin{array}{cc}\bar{y} & -x \\ \bar{x} & -y  \end{array} & \begin{array}{cc}\frac{1-a}{2} & \bar{t} \\ t & \frac{1-a}{2} \end{array}       \end{array} \right ]  \label{b6}
\end{equation}
with rank 2. By Proposition 2.1 this implies that
\begin{equation}
\left [\begin{array}{cc} \frac{1-a}{2} & \bar{t} \\ t & \frac{1-a}{2}  \end{array} \right ] = A^{\dagger} KA \label{b7}
\end{equation}
where
\begin{equation}
A = K^{-1} \left [\begin{array}{cc}y & x \\ - \bar{x} & - \bar{y}    \end{array}   \right ], \quad K = \left [\begin{array}{cc}\frac{a}{2} & z \\ \bar{z} & \frac{a}{2}  \end{array} \right ]  \label{b8}
\end{equation}
Putting $x = \frac{\sqrt{a(1-a)}}{2} e^{i \theta},$ $y = \frac{\sqrt{a(1-a)}}{2} e^{i \varphi},$ substituting the expressions of (\ref{b8}) in (\ref{b7}) and equating the 11-entry of the matrices on both sides of (3.7)  we get
$$\left |\frac{a}{2} + z \: e^{-i (\theta + \varphi)}  \right |^2 = 0  $$
and therefore $|z|^2 = \frac{a^2}{4},$ a contradiction.

The case $|t|^2 < \frac{(1-a)^2}{4}$ is dealt with in the same manner.

Now we shall prove that $\rho$ is not extremal. Express (\ref{b4}) as
\begin{equation}
K_{\rho} = \left [\begin{array}{c|c} K & KA \\ \hline A^{\dagger} K & A^{\dagger} KA   \end{array}  \right ] \label{b9}
\end{equation}
where
\begin{equation}
K = \left [\begin{array}{cc} \frac{a}{2} & x \\ \bar{x} & \frac{1-a}{2}  \end{array}  \right ], \quad A = K^{-1} \left [\begin{array}{cc}y & z \\ t & -y  \end{array}  \right ]   \label {b10}
\end{equation}

\begin{equation}
A^{\dagger} KA = d K^{-1}, \quad d = \frac{a(1-a)}{4} - |x|^2 > 0 \label {b11}
\end{equation}
This implies the existence of a unitary matrix $U$ such that 
$$K^{\frac{1}{2}} A = d^{\frac{1}{2}} U K ^{- \frac{1}{2}} .$$
From (\ref{b10}) we have
$$ \left [ \begin{array}{cc}y & z \\ t & -y  \end{array} \right ] = KA = d^{1/2} K^{1/2} U K^{-1/2}.$$
Hence $\mbox{Tr}\: U = 0.$ Since $U$ is a unitary matrix of zero trace it has the form
$$U = e^{i \theta} \: V $$
where $V$ is a selfadjoint  unitary matrix of determinant $-1.$ In particular
\begin{equation}
A = d^{1/2} e^{i \theta} K^{-1/2} V K^{-1/2} \label{b12}
\end{equation}
where $V$ is selfadjoint and unitary. We now examine the linear space
\begin{equation}
\cld = \left \{ \left .[I_2 | A] \left (X_1 \otimes I_2 + I_2 \otimes X_2 \right ) [\frac{I_2}{A^{t}}] \right | X_i \: \mbox{is hermitian for each}\: i    \right \}. \label{b13}
\end{equation}
In the ordered basis $\{\bbe_j,  \:\:j=1,2,3,4\}$ it is easily verified that $X_1 \otimes I_2 + I_2 \otimes X_2$ in $\cld$ varies over all matrices of the form
$$ \left \{ \left . \left [ \begin{array}{c|c}X + p I_2 & r I_2 \\ \hline \bar{r} I_2 & X + q I_2    \end{array}  \right ] \right | X \:\mbox{hermitian,} \: p, q \in  \IR, r \in \IC   \right \}.$$
Thus
$$\cld = \left \{\left . X + AXA^{\dagger} + rA^{\dagger} + \bar{r}A + qAA^{\dagger} + pI \right | X \: \:\mbox{hermitian,} \: p, q \varepsilon \IR, r \in \IC  \right \}.$$
We now search for a hermitian matrix $L$ of order 2 in $\cld^{\perp}$ with respect to the scalar product $\langle X_1| X_2 \rangle = \mbox{Tr}\: X_1 X_2$ for any two hermitian matrices of order 2. In other words we search for a hermitian $L$ satisfying
\begin{equation}
\left . \begin{array}{l}  \mbox{Tr}\:\: L  = 0, \:\: \mbox{Tr} \:\: L \: K^{-1/2} V \:K^{1/2} = 0 \\ 
\mbox{Tr}\:\: L \left (X + d K^{-1/2} V K^{-1/2} X  K^{-1/2} V K^{-1/2} \right ) = 0  \end{array} \right \}  \label{b14}
\end{equation}
for all  hermitian $X.$ (Here we have substituted for $A$ from (\ref{b12})).

Note that $\sqrt{d} K^{-1/2} V K^{-1/2} = B$ is a hermitian matrix of determinant $-1.$ Thus (\ref{b14}) reduces to
\begin{equation}
\mbox{Tr}\:\: L = 0, \quad \mbox{Tr}\:\: LB = 0, \quad L + BLB = 0.   \label{b15}
\end{equation}
The matrix $B$ can be expressed as
$$B = WDW^{t} $$
where $W$ is unitary and
$$D = \left [\begin{array}{cc}\alpha & 0 \\ 0 & - \alpha^{-1} \end{array} \right ], \quad \alpha > 0. $$
Then for any $\xi \in \IC$ the hermitian matrix
$$L = W^{t} \left [\begin{array}{cc}0 & \xi \\ \bar{\xi} & 0   \end{array}   \right ] W$$
satisfies (\ref{b15}). In other words $\cld^{\perp} \neq \{0\}$ and therefore the linear space $\cld$ in (\ref{b13}) is not the space of all hermitian matrices of order 2. Hence by Theorem 2.5, the state $\rho$ is not extremal.

Thus every extremal state $\rho$ in $\clc (\frac{1}{2}I, \frac{1}{2}I)$ is of rank 1. Such an extremal state $\rho$ has the form
$$\rho = | \Omega >< \Omega |$$
where
\begin{eqnarray*}
| \Omega >& =& \sum_{x,y \in \{0,1\}}    a_{xy} | xy>,  \\
&& \sum_{x,y} |a_{xy}|^2 = 1.
\end{eqnarray*}
The fact that $| \Omega >< \Omega |$ has its marginal operators equal to $\frac{1}{2} I$ implies that $((a_{xy})) = \frac{1}{\sqrt{2}} ((u_{xy}))$ where $((u_{xy}))$ is a unitary matrix of order 2. Putting
$$\sum_{y=0}^{1} u _{xy} | y > = | \psi_{x} > $$
we see that
\begin{equation}
| \Omega > = \frac{1}{\sqrt{2}} \left (|0> | \psi_0> + |1> |\psi_1 > \right )   \label{b16}
\end{equation}
where $\{|0>, |1>  \}$ is the canonical orthonormal basis in $\IC^2$ and $\{|\psi _0 >, | \psi_1 >\}$ is another orthonormal basis in $\IC^2$ (which may coincide with $\{|0>, |1>\}$). Varying the orthonormal basis $\{|\psi_0>, |\psi_1>\}$ of $\IC^2$ in (\ref{b16}) we get all the extremal states of $\clc (\frac{1}{2} I, \frac{1}{2}I)$ as $|\Omega ><\Omega|.$ \qed

\vskip1in
\noindent {\bf References}
\vskip20pt
\begin{enumerate}
\item G. Birkhoff, {\it Three observations on linear algebra,} Univ. Nac. Tucum\'an Rev. Ser. A5 (1946) 147-151.

\item J. P. R. Christensen and J. Vesterstrom, {\it A note on extreme positive definite matrices,} Math. Ann. 244 (1979) 65-68.

\item R. Grome, S. Pierce and W. Watkins, {\it Extremal correlation matrices,} Lin. alg. and Appl. 134 (1990) 63-70.

\item R. Loewy, {\it Extreme points of a convex subset of the cone of positive semidefinite matrices, } Math. Ann. 253 (1980) 227-232.

\item K. R. Parthasarathy, {\it On extremal correlations,} J. Stat. Planning and Inf. 103 (2002) 73-80.
\end{enumerate}

\end{document}